\documentclass[pra,fleqn,floatfix]{revtex4}

\usepackage{amsmath}
\usepackage{dcolumn}

\begin{document}

\title{Leading order relativistic corrections to the ro-vibrational spectrum of $\mbox{H}^+_2$ and $\mbox{HD}^+$ molecular ions}

\author{D.T.~Aznabayev$^{1,2,3}$}
\author{A.K.~Bekbaev$^{1,4}$}
\author{Vladimir I. Korobov$^{1,5}$}
\affiliation{$^1$Bogoliubov Laboratory of Theoretical Physics, Joint Institute
for Nuclear Research, Dubna 141980, Russia,}
\affiliation{$^2$The Institute of Nuclear Physics, Ministry of Energy of the Republic of Kazakhstan, 050032 Almaty, Kazakhstan}
\affiliation{$^3$L.N.~Gumilyov Eurasian National University, 010000 Astana, Kazakhstan}
\affiliation{$^4$Al-Farabi Kazakh National University, 050038 Almaty, Kazakhstan}
\affiliation{$^5$Peoples' Friendship University of Russia (RUDN University), 6 Miklukho-Maklaya St, Moscow, 117198, Russia}

\begin{abstract}
High-precision variational calculations of the operators for the relativistic corrections in the leading $m\alpha^4$ order are presented. The ro-vibrational states in the range of the total orbital angular momentum $L\!=\!0\!-\!4$ and vibrational quantum number $v\!=\!0\!-\!10$ for the $\mbox{H}_2^+$ and $\mbox{HD}^+$ molecular ions are considered. We estimate that about ten significant digits are obtained. This high precision is required for making theoretical predictions for transition frequencies at the level of $10^{-12}$ relative uncertainty.
\end{abstract}

\maketitle

In recent years several laser experiments to measure vibrational transitions with high precision in the hydrogen molecular ions $\mbox{HD}^+$ have been performed \cite{expPRL07,Schiller12,exp16}. One of the trends of these and new yet planned experiments \cite{KarrPatra16} is substantial increase of the vibrational number of the final state from $v=1$ to $v=8$ or 9. That makes necessary new theoretical calculations, which covers states with high $v$, in order to comply with the requirements of the experiments.

Recently a very promising pure rotational transition experiment has been carried out \cite{SchillerNature18}, which realized conditions of the Lamb-Dicke regime and got for the first time experimental value of the transition frequency with relative precision of $3\cdot10^{-10}$. It is realistic that this precision may be improved in a recent future by about two orders of magnitude. That gives another challenge to theory \cite{KorPRL17}, since calculation of theoretical frequencies for pure vibrational transitions leads to strong cancelations thus require very high accuracy in calculation of the matrix elements of various contributions, at least those, which are connected with leading order relativistic corrections.

The variational calculations of the nonrelativistic energies during past years have reached a numerical precision of $10^{-15}-10^{-30}$ a.u.\ \cite{schiller-korobov,Drake04,LiYan07,Nakatsuji09}. The ultimate accuracy of $\sim10^{-34}$ a.u.\ has been obtained for the $\mbox{H}_2^+$ molecular ion ground state \cite{NingYan14}. These calculations demonstrate that at least the nonrelativistic ro-vibrational transition frequencies can be determined with the accuracy well below the 1 Hz level.

So far, systematic calculation of the leading order relativistic and radiative corrections were obtained in \cite{KorPRA06,ZhongPRA09,ZhongPRA12,ZhongCPB15}. In all the cases the vibration quantum number has been restricted to $v=4$. It is worth noting that in the last publication, precision of the matrix elements for the Breit-Pauli matrix elements for the ground vibration states, $v=0$, are very high and definitely sufficient for pure rotation transition case. But with increase of $v$ numerical precision of the calculations based on the Hylleraas variational expansion drops down very rapidly. So, these challenges requires further numerical efforts to extend and to make more precise data on the leading order relativistic corrections.

The next important step is evaluation of the relativistic and radiative corrections to the binding energies of the ro-vibrational levels. This can be systematically performed using series expansion of the binding energy
in terms of the coupling constant, in our case, the fine structure constant, $\alpha$. The key quantity for the leading order $R_\infty\alpha^3$ radiative correction, the Bethe logarithm, for $\mbox{HD}^+$ and $\mbox{H}_2^+$ have been obtain in our previous work \cite{BL_PRA12}. This work is aimed to extend calculations of the leading order relativistic corrections to the larger range of vibrational states and to improve precision of the mean values of the operators for the high $v$ states.

The following notation is used throughout this paper. $\mathbf{P}_1$, $\mathbf{P}_2$, and $\mathbf{p}_e$ are the momenta and $\mathbf{R}_1$, $\mathbf{R}_2$, $\mathbf{r}_e$ are the coordinates of nuclei and electron with respect to the center of mass of a molecule, and
\[
\mathbf{r}_1=\mathbf{r}_e-\mathbf{R}_1,
\quad
\mathbf{r}_2=\mathbf{r}_e-\mathbf{R}_2,
\quad
\mathbf{R}=\mathbf{R}_2-\mathbf{R}_1.
\]
Here we assume that indices 1 and 2 stand for the protons in case of $\mbox{H}_2^+$, and $\mathbf{R}_1\!\equiv\!\mathbf{R}_d$ --- the coordinate of a deuteron in case of $\mbox{HD}^+$. The atomic units ($\hbar=e=m_e=1$) are employed. We use the CODATA14 recommended values of the fundamental constants \cite{CODATA14} for all our calculations.

\section{Variational wave function}

\begin{table}[b]
\begin{center}
\caption{Convergence of the mean value of $\left\langle\mathbf{p}_e^4\right\rangle$ for the vibrational states $v=4$ and $v=10$ in $\mbox{HD}^+$ for various $L$.} \label{converg}
\begin{tabular}{l@{\hspace{2mm}}r@{\hspace{5mm}}l@{\hspace{5mm}}l}
\hline\hline
\vrule width0pt height 11pt
 & $N$~~ & ~~~~$L=0$ & ~~~~$L=3$ \\
\hline
\vrule width0pt height 11pt
$v=4$ &  8\,000   & 5.792\,077\,3804   & 5.765\,200\,6482 \\
      & 10\,000   & 5.792\,077\,3799   & 5.765\,200\,6473 \\
      & 12\,000   & 5.792\,077\,3798   & 5.765\,200\,6468 \\
      & 14\,000   & 5.792\,077\,3798   & 5.765\,200\,6466 \\
      & $\infty$~~& 5.792\,077\,3798(1)& 5.765\,200\,6465(1)\\[1mm]
&Ref.~\cite{ZhongPRA12}
      & 5.792\,077\,379\,5339(5) & 5.765\,200\,6889(6) \\
\hline
$v=10$& 18\,000   & 5.276\,903\,8407   & 5.259\,382\,5408 \\
      & 20\,000   & 5.276\,903\,8401   & 5.259\,382\,5364 \\
      & 22\,000   & 5.276\,903\,8397   & 5.259\,382\,5339 \\
      & 24\,000   & 5.276\,903\,8395   & 5.259\,382\,5328 \\
      & $\infty$~~& 5.276\,903\,8394(2)& 5.259\,382\,5322(9) \\
\hline\hline
\end{tabular}
\end{center}
\end{table}

The variational bound state wave functions were calculated by solving the three-body Schr\"{o}dinger equation with Coulomb interaction:
\begin{equation}
\left[
   \frac{\mathbf{P}_1^2}{2M_1}+\frac{\mathbf{P}_2^2}{2M_2}+\frac{\mathbf{p}_e^2}{2m}
      -\frac{Z_1}{r_1}-\frac{Z_2}{r_2}+\frac{Z_1Z_2}{R}
\right]\Psi_0 = E_0\Psi_0,
\end{equation}
using the variational approach based on the exponential expansion with randomly chosen exponents. This approach has been discussed and developed in a variety of works \cite{early,Thakkar,Frolov95}. Details and particular strategy of choice of the variational nonlinear parameters and basis structure that have been adopted in the present work can be found in \cite{Kor00}.

Briefly, the wave function for a state with a total orbital angular
momentum $L$ and of a total spatial parity $\pi=(-1)^L$ is expanded as
follows:
\begin{equation}\label{exp_main}
\begin{array}{@{}l}
\displaystyle \Psi_{LM}^\pi(\mathbf{R},\mathbf{r}_1) =
       \sum_{l_1+l_2=L}
         \mathcal{Y}^{l_1l_2}_{LM}(\hat{\mathbf{R}},\hat{\mathbf{r}}_1)
         G^{L\pi}_{l_1l_2}(R,r_1,r_2),
\\[4mm]\displaystyle
G_{l_1l_2}^{L\pi}(R,r_1,r_2) =
    \sum_{n=1}^N \Big\{C_n\,\mbox{Re}
          \bigl[e^{-\alpha_n R-\beta_n r_1-\gamma_n r_2}\bigr]
+D_n\,\mbox{Im} \bigl[e^{-\alpha_n R-\beta_n r_1-\gamma_n r_2}\bigr] \Big\}.
\end{array}
\end{equation}
where the complex exponents, $\alpha$, $\beta$, $\gamma$, are generated in
a pseudorandom way.

When exponents $\alpha_n$, $\beta_n$, and $\gamma_n$ are real, the method reveals slow convergence for molecular type Coulomb systems. Thus the use of complex exponents allows to reproduce the oscillatory behaviour of the vibrational part of the wave function and to improve convergence \cite{Frolov95,Kor00}.

\begin{table}
\caption{Mean values of various operators for the ro-vibrational states $(L,v)$ in the H$_2^+$ molecular ion.}
\label{H2plus-av}
\begin{center}
\begin{tabular}{cr@{\hspace{4mm}}c@{\hspace{4mm}}c@{\hspace{4mm}}c@{\hspace{4mm}}c@{\hspace{4mm}}c@{\hspace{4mm}}c}
\hline\hline
\vrule width0pt height 11pt
 & $v$ & $\langle\textbf{p}^4_e\rangle$ & $\langle\delta(\textbf{r}_1)\rangle$ & $\langle\textbf{P}^4_1\rangle$ &
$R_{pe}$ &$R_{pp}$ &$Q_{pe}$ \\
\hline
\vrule width0pt height 12pt
$L=0$& 0&6.2856600594   &0.20673647629  &79.797649364   &1.170117625    &4.601934314    &$-$0.1344262279 \\
  & 1   &6.1245198079   &0.20131066471  &334.89830216   &1.140805227    &12.89614650    &$-$0.1312863754 \\
  & 2   &5.9762285601   &0.19629458830  &762.80369623   &1.114077226    &19.87898102    &$-$0.1283925753 \\
  & 3   &5.8400118491   &0.19166249726  &1304.2092864   &1.089801609    &25.66162135    &$-$0.1257315001 \\
  & 4   &5.7151984964   &0.18739184669  &1908.3716822   &1.067868274    &30.33829429    &$-$0.1232916813 \\
  & 5   &5.6012112574   &0.18346299199  &2531.6173039   &1.048187719    &33.98804802    &$-$0.1210633708 \\
  & 6   &5.4975592498   &0.17985893910  &3136.1167480   &1.030690164    &36.67618495    &$-$0.1190384366 \\
  & 7   &5.4038319567   &0.17656514348  &3688.8844922   &1.015325087    &38.45539351    &$-$0.1172102897 \\
  & 8   &5.3196946603   &0.17356935365  &4160.9753024   &1.002061184    &39.36661303    &$-$0.1155738424 \\
  & 9   &5.2448852254   &0.17086149709  &4526.8597586   &0.9908868010   &39.43965837    &$-$0.1141255007 \\
  & 10  &5.1792121777   &0.16843360820  &4763.9728538   &0.9818108626   &38.69362778    &$-$0.1128631936 \\[1mm]
$L=1$& 0&6.2780390374   &0.20649132016  &85.050455667   &1.168818664    &4.834336475    &$-$0.1342668169 \\
  & 1   &6.1173973078   &0.20108117308  &347.54872567   &1.139597149    &13.09529203    &$-$0.1311369254 \\
  & 2   &5.9695793176   &0.19607994205  &780.92161614   &1.112955682    &20.04770789    &$-$0.1282526561 \\
  & 3   &5.8338140029   &0.19146198172  &1326.1219910   &1.088762926    &25.80234570    &$-$0.1256007425 \\
  & 4   &5.7094333595   &0.18720484414  &1932.6198442   &1.066909426    &30.45305169    &$-$0.1231697745 \\
  & 5   &5.5958631574   &0.18328897611  &2556.9179839   &1.047306308    &34.07852553    &$-$0.1209500612 \\
  & 6   &5.4926153921   &0.17969747034  &3161.3322085   &1.029884418    &36.74374470    &$-$0.1189335254 \\
  & 7   &5.3992823972   &0.17641586729  &3712.9958575   &1.014593872    &38.50108793    &$-$0.1171136355 \\
  & 8   &5.3155323072   &0.17343199956  &4183.0603098   &1.001404040    &39.39119123    &$-$0.1154853623 \\
  & 9   &5.2411059278   &0.17073588027  &4546.0740115   &0.9903039968   &39.44356252    &$-$0.1140451747 \\
  & 10  &5.1758148732   &0.16831963269  &4779.5341125   &0.9813034913   &38.67697691    &$-$0.1127910691 \\[1mm]
$L=2$& 0&6.2629099578   &0.20600454283  &96.910905201   &1.166239007    &5.293427390    &$-$0.1339504598 \\
  & 1   &6.1032596294   &0.20062554856  &373.99761897   &1.137198302    &13.48828656    &$-$0.1308403800 \\
  & 2   &5.9563828602   &0.19565384281  &818.12463975   &1.110729011    &20.38024874    &$-$0.1279750672 \\
  & 3   &5.8215153662   &0.19106398834  &1370.7553673   &1.086701150    &26.07923567    &$-$0.1253413771 \\
  & 4   &5.6979954729   &0.18683373048  &1981.7828915   &1.065006550    &30.67833547    &$-$0.1229280161 \\
  & 5   &5.5852549395   &0.18294369775  &2608.0596187   &1.045557594    &34.25555325    &$-$0.1207254060 \\
  & 6   &5.4828115576   &0.17937715716  &3212.1889678   &1.028286377    &36.87521850    &$-$0.1187255828 \\
  & 7   &5.3902633187   &0.17611981800  &3761.5395264   &1.013144294    &38.58909168    &$-$0.1169221266 \\
  & 8   &5.3072840707   &0.17315968126  &4227.4536463   &1.000102065    &39.43720246    &$-$0.1153101265 \\
  & 9   &5.2336204946   &0.17048693110  &4584.6333862   &0.9891502316   &39.44844397    &$-$0.1138861763 \\
  & 10  &5.1690904704   &0.16809387031  &4810.6981087   &0.9803002006   &38.64094245    &$-$0.1126484120 \\[1mm]
$L=3$& 0&6.2404943312   &0.20528307706  &117.97635160   &1.162414502    &5.968061524    &$-$0.1334819874 \\
  & 1   &6.0823171617   &0.19995037649  &416.43490453   &1.133642648    &14.06480756    &$-$0.1304013519 \\
  & 2   &5.9368391490   &0.19502254993  &876.24662762   &1.107429423    &20.86703769    &$-$0.1275642145 \\
  & 3   &5.8033061164   &0.19047447029  &1439.6301087   &1.083646859    &26.48342299    &$-$0.1249576135 \\
  & 4   &5.6810658304   &0.18628417107  &2057.1039760   &1.062188713    &31.00592204    &$-$0.1225704294 \\
  & 5   &5.5695590038   &0.18243255228  &2686.0371498   &1.042969244    &34.51150425    &$-$0.1203932534 \\
  & 6   &5.4683120665   &0.17890314079  &3289.4575739   &1.025922422    &37.06353165    &$-$0.1184182891 \\
  & 7   &5.3769315140   &0.17568189914  &3835.0805572   &1.011001564    &38.71284003    &$-$0.1166392866 \\
  & 8   &5.2950997615   &0.17275708064  &4294.5293028   &0.9981794217   &39.49855014    &$-$0.1150515119 \\
  & 9   &5.2225723483   &0.17011912691  &4642.7309921   &0.9874487559   &39.44863056    &$-$0.1136517483 \\
  & 10  &5.1591765901   &0.16776061086  &4857.4830325   &0.9788234896   &38.58022733    &$-$0.1124383412 \\[1mm]
$L=4$& 0&6.2111137341   &0.20433699022  &151.87439790   &1.157397265    &6.842198150    &$-$0.1328684048 \\
  & 1   &6.0548751988   &0.19906521543  &477.89237049   &1.128979542    &14.81002209    &$-$0.1298265267 \\
  & 2   &5.9112382325   &0.19419514510  &957.79934933   &1.103103711    &21.49435017    &$-$0.1270264811 \\
  & 3   &5.7794620170   &0.18970206193  &1534.8012769   &1.079644470    &27.00220305    &$-$0.1244555480 \\
  & 4   &5.6589068100   &0.18556438053  &2160.2340467   &1.058498128    &31.42404737    &$-$0.1221028394 \\
  & 5   &5.5490251876   &0.18176336040  &2792.1407251   &1.039581440    &34.83548285    &$-$0.1199591710 \\
  & 6   &5.4493550513   &0.17828287309  &3394.1025379   &1.022830872    &37.29858981    &$-$0.1180169703 \\
  & 7   &5.3595142132   &0.17510921788  &3934.2857135   &0.1008202311   &38.86297608    &$-$0.1162702141 \\
  & 8   &5.2791965306   &0.17223098447  &4384.6775195   &0.9956712337   &39.56655122    &$-$0.1147144056 \\
  & 9   &5.2081694237   &0.16963895903  &4720.4954258   &0.9852333961   &39.43604596    &$-$0.1133465831 \\
  & 10  &5.1462729334   &0.16732607784  &4919.7643085   &0.9769061357   &38.48728563    &$-$0.1121653755 \\
\hline\hline
\end{tabular}
\end{center}
\end{table}

\section{Leading order relativistic corrections}

The leading order relativistic corrections ($R_\infty\alpha^2$) at present are well understood and are described by the Breit-Pauli Hamiltonian. Consideration of this part can be found in many textbooks, see, for example, Refs.~\cite{BS,BLP}, or a comprehensive review \cite{Eides01}. Here we present in explicit form expressions for different terms, which contribute to this order.

The major contribution comes from the relativistic correction for the bound electron,
\begin{equation}\label{rc}
E_{rc}^{(2)} = \alpha^2\!\left\langle\!
         -\frac{\mathbf{p}_e^4}{8m_e^3}
         +\frac{4\pi }{8m_e^2}
           \left[Z_1\delta(\mathbf{r}_1)
                +Z_2\delta(\mathbf{r}_2)
           \right]
      \right\rangle.
\end{equation}

The other corrections are due to a finite mass of nuclei and are called the recoil corrections of orders $R_\infty\alpha^2(m/M)$, $R_\infty\alpha^2(m/M)^2$, etc. The first is the transverse photon exchange,
\begin{equation}\label{trans}
\begin{array}{@{}l}
\displaystyle
E_{tr\text{-}ph}^{(2)} =
\frac{\alpha^2Z_1}{2m_eM_1}
  \left\langle
     \frac{\mathbf{p}_e\mathbf{P}_1}{r_1}
    +\frac{\mathbf{r}_1(\mathbf{r}_1\mathbf{p}_e)\mathbf{P}_1}{r_1^3}
  \right\rangle+
\frac{\alpha^2Z_2}{2m_eM_2}
  \left\langle
     \frac{\mathbf{p}_e\mathbf{P}_2}{r_2}
    +\frac{\mathbf{r}_2(\mathbf{r}_2\mathbf{p}_e)\mathbf{P}_2}{r_2^3}
  \right\rangle
\\[4mm]\displaystyle\hspace{30mm}
-\frac{\alpha^2Z_1Z_2}{2M_1M_2}
  \left\langle
     \frac{\mathbf{P}_1\mathbf{P}_2}{R}
    +\frac{\mathbf{R}(\mathbf{R}\mathbf{P}_1)\mathbf{P}_2}{R^3}
  \right\rangle.
\end{array}
\end{equation}
The contribution of the last term in (\ref{trans}) is not negligible and amounts to about 10\% of $E_{tr\text{-}ph}^{(2)}$.

The next is the relativistic kinetic energy ($E_{kin}=\sqrt{m^2\!+\!p^2}\approx m\!+\!p^2/2m\!+\!\dots$) correction for heavy particles,
\begin{equation}
E_{kin}^{(2)} = -\alpha^2\left\langle
              \frac{\mathbf{P}_1^4}{8M_1^3}
              +\frac{\mathbf{P}_2^4}{8M_2^3}
       \right\rangle.
\end{equation}

Further in the $R_\infty\alpha^2$ order, one has to consider the nuclear spin dependent recoil corrections. For the proton, spin 1/2 particle, one has
\begin{equation}
E_{\rm Darwin}^{(2)} = \frac{\alpha^24\pi Z_p}{8M_p^2}
      \bigl\langle
         \delta(\mathbf{r}_p)
      \bigr\rangle.
\end{equation}
In case of deuteron (spin one) this term vanishes. The leading order electric charge finite size correction is defined (both for proton and deuteron) by
\begin{equation}
E_{\rm nuc}^{(2)} = \sum_{i=1,2}
 \frac{2\pi Z_i(R_i/a_0)^2}{3} \bigl\langle \delta(\mathbf{r}_i) \bigr\rangle,
\end{equation}
where $R$ is the root-mean-square radius of the nuclear electric charge distribution. The RMS radius for the proton is: $R_p = 0.8750(68)\mbox{ fm}$, and for the deuteron: $R_d = 2.1394(28)\mbox{ fm}$. These contributions are connected with internal structure of complex particles. For a detailed discussion of this rather nontrivial problem we refer the reader to \cite{Pachucki94,Khr96}.

The complete contribution to this order thus is:
\begin{equation}
E_{\alpha^2} =
   E_{rc}^{(2)}+E_{kin}^{(2)}+E_{tr\text{-}ph}^{(2)}
   +E_{\rm Darwin}^{(2)}+E_{\rm nuc}^{(2)}.
\end{equation}

\begin{table}[t]
\caption{Mean values of operators for HD$^{+}$.}
\label{HDplus-av}
\begin{center}
\begin{tabular}{cr@{\hspace{4mm}}c@{\hspace{4mm}}c@{\hspace{4mm}}c@{\hspace{4mm}}c@{\hspace{4mm}}c}
\hline\hline
\vrule width0pt height 11pt
& $v$ & $\left\langle\delta(\textbf{r}_d)\right\rangle$ & $\left\langle\delta(\textbf{r}_p)\right\rangle$ & $\left\langle\mathbf{p}_e^{4}\right\rangle$ & $\left\langle\mathbf{P}_p^{4}\right\rangle$ & $\left\langle\mathbf{P}_d^{4}\right\rangle$ \\
\hline
\vrule width0pt height 12pt
$L=0$
&$0$ &0.20734814178  &0.20704259948  &6.3001999477   &104.37171376   &104.44384898 \\
&$1$ &0.20260117861  &0.20228886474  &6.1590223524   &449.45675982   &449.73914662 \\
&$2$ &0.19816679513  &0.19784583748  &6.0276143229   &1042.8168322   &1043.4433174 \\
&$3$ &0.19402784139  &0.19369601393  &5.9054534327   &1812.9975669   &1814.0632379 \\
&$4$ &0.19016909090  &0.18982370286  &5.7920773798   &2697.6692759   &2699.2354173 \\
&$5$ &0.18657710159  &0.18621484444  &5.6870790520   &3642.2321785   &3644.3304724 \\
&$6$ &0.18324011137  &0.18285684358  &5.5901023863   &4598.6380478   &4601.2741340 \\
&$7$ &0.18014797184  &0.17973840936  &5.5008389477   &5524.3976372   &5527.5541366 \\
&$8$ &0.17729212821  &0.17684938788  &5.4190251466   &6381.7503195   &6385.3894083 \\
&$9$ &0.17466566245  &0.17418056963  &5.3444400643   &7136.9787737   &7141.0443898 \\
&$10$&0.17226343214  &0.17172343636  &5.2769038394   &7759.8577177   &7764.2774745 \\[1mm]
$L=1$
&$0$ &0.20716324168  &0.20685769957  &6.2944507461   &110.38493989   &110.46133144 \\
&$1$ &0.20242655717  &0.20211421221  &6.1535997460   &464.36193851   &464.65338756 \\
&$2$ &0.19800198124  &0.19768095602  &6.0225036793   &1064.6241336   &1065.2634075 \\
&$3$ &0.19387241680  &0.19354047839  &5.9006418101   &1839.9516894   &1841.0329235 \\
&$4$ &0.19002268724  &0.18967713562  &5.7875534291   &2728.2144184   &2729.7980540 \\
&$5$ &0.18643939814  &0.18607691220  &5.6828329367   &3674.9819497   &3677.0989179 \\
&$6$ &0.18311083326  &0.18272725426  &5.5861257049   &4632.3495498   &4635.0048177 \\
&$7$ &0.18002688970  &0.17961690940  &5.4971246933   &5557.9490993   &5561.1246812 \\
&$8$ &0.17717905975  &0.17673576024  &5.4155677242   &6414.1217749   &6417.7792967 \\
&$9$ &0.17456047384  &0.17407462955  &5.3412352361   &7167.2353693   &7171.3182658 \\
&$10$&0.17216604459  &0.17162502783  &5.2739488088   &7787.1353273   &7791.5707486 \\[1mm]
$L=2$
&$0$ &0.20679544434  &0.20648990098  &6.2830163299   &123.79001421   &123.87565882 \\
&$1$ &0.20207923036  &0.20176682151  &6.1428157396   &495.36790199   &495.67811676 \\
&$2$ &0.19767418733  &0.19735302514  &6.0123409636   &1109.2715320   &1109.9369362 \\
&$3$ &0.19356332376  &0.19323116142  &5.8910746483   &1894.7472178   &1895.8600523 \\
&$4$ &0.18973156234  &0.18938568123  &5.7785592512   &2790.0612338   &2791.6802614 \\
&$5$ &0.18616560334  &0.18580265714  &5.6743921721   &3741.1197981   &3743.2744544 \\
&$6$ &0.18285382284  &0.18246961801  &5.5782216799   &4700.3031879   &4702.9971004 \\
&$7$ &0.17978620914  &0.17937538868  &5.4897435379   &5625.4837721   &5628.6977473 \\
&$8$ &0.17695434714  &0.17650992330  &5.4086983151   &6479.2047371   &6482.8993039 \\
&$9$ &0.17435146623  &0.17386411066  &5.3348692560   &7228.0026311   &7232.1202214 \\
&$10$&0.17197258810  &0.17142951724  &5.2680807597   &7841.8629641   &7846.3298040 \\[1mm]
$L=3$
&$0$ &0.20624869702  &0.20594314769  &6.2660227900   &147.25876390   &147.36009301 \\
&$1$ &0.20156297374  &0.20125046479  &6.1267909226   &544.78450154   &545.12442395 \\
&$2$ &0.19718702656  &0.19686565434  &5.9972415658   &1178.7470290   &1179.4529692 \\
&$3$ &0.19310402154  &0.19277151828  &5.8768624098   &1979.0846625   &1980.2460433 \\
&$4$ &0.18929902981  &0.18895264869  &5.7652006466   &2884.6520623   &2886.3251483 \\
&$5$ &0.18575889338  &0.18539524983  &5.6618581670   &3841.8547847   &3844.0667859 \\
&$6$ &0.18247212603  &0.18208697377  &5.5664875531   &4803.4957133   &4806.2482620 \\
&$7$ &0.17942885267  &0.17901676065  &5.4787887613   &5727.8035584   &5731.0756611 \\
&$8$ &0.17662079781  &0.17617467198  &5.3985065094   &6577.6205427   &6581.3710932 \\
&$9$ &0.17404133892  &0.17355169431  &5.3254282095   &7319.7326808   &7323.9026143 \\
&$10$&0.17168566289  &0.17113947797  &5.2593825328   &7924.3321151   &7928.8462782 \\[1mm]
$L=4$
&$0$ &0.20552877799  &0.20522321313  &6.2436546662   &184.59247486   &184.71796048 \\
&$1$ &0.20088331252  &0.20057066224  &6.1057017312   &615.88018454   &616.26250830 \\
&$2$ &0.19654578490  &0.19622412424  &5.9773742631   &1275.8456943   &1276.6080718 \\
&$3$ &0.19249956988  &0.19216660270  &5.8581666316   &2095.3366407   &2096.5647803 \\
&$4$ &0.18872993534  &0.18838287675  &5.7476323025   &3013.9805344   &3015.7274073 \\
&$5$ &0.18522391134  &0.18485932503  &5.6453791232   &3978.8384624   &3981.1283423 \\
&$6$ &0.18197019460  &0.18158376287  &5.5510654021   &4943.2676976   &4946.0995834 \\
&$7$ &0.17895909192  &0.17854528318  &5.4643966701   &5865.9639410   &5869.3144615 \\
&$8$ &0.17618251410  &0.17573408983  &5.3851231705   &6710.1607564   &6713.9866433 \\
&$9$ &0.17363403687  &0.17314129869  &5.3130378763   &7442.9699688   &7447.2101770 \\
&$10$ &0.17130906936 &0.17075867070  &5.2479751736   &8034.8527748   &8039.4303232 \\
\hline\hline
\end{tabular}
\end{center}
\end{table}

\begin{table}[t]\addtocounter{table}{-1}
\caption{(continued) Mean values of operators for HD$^{+}$.}
\begin{center}
\begin{tabular}{cr@{\hspace{4mm}}c@{\hspace{4mm}}c@{\hspace{4mm}}c@{\hspace{4mm}}c@{\hspace{4mm}}c}
\hline\hline
\vrule width0pt height 11pt
&$v$     & $R_{dp}$     &$ R_{de}$     & $R_{pe}$      &$Q(r_{1})$     &$Q(r_{12})$ \\
\hline
\vrule width0pt height 12pt
$L=0$
&$0$ &5.354630521    &1.174487826    &1.170770145    &$-$0.1348622766    &$-$0.1345911956 \\
&$1$ &15.14482590    &1.150481264    &1.143366405    &$-$0.1321133551    &$-$0.1318389778 \\
&$2$ &23.59899698    &1.128236201    &1.118123569    &$-$0.1295519876    &$-$0.1292729300 \\
&$3$ &30.81765302    &1.107677206    &1.094937543    &$-$0.1271692905    &$-$0.1268839283 \\
&$4$ &36.88807889    &1.088739930    &1.073718404    &$-$0.1249574956    &$-$0.1246638869 \\
&$5$ &41.88558857    &1.071370643    &1.054389416    &$-$0.1229098904    &$-$0.1226056689 \\
&$6$ &45.87457090    &1.055526002    &1.036886189    &$-$0.1210207797    &$-$0.1207030052 \\
&$7$ &48.90935198    &1.041173055    &1.021155946    &$-$0.1192854727    &$-$0.1189504164 \\
&$8$ &51.03489389    &1.028289564    &1.007156830    &$-$0.1177003010    &$-$0.1173431284 \\
&$9$ &52.28734536    &1.016864765    &0.9948571418   &$-$0.1162626814    &$-$0.1158769724 \\
&$10$&52.69445717    &1.006900770    &0.9842343052   &$-$0.1149712469    &$-$0.1145482435 \\[1mm]
$L=1$
&$0$ &5.589562257    &1.173578500    &1.169719380    &$-$0.1347420011    &$-$0.1344709788 \\
&$1$ &15.35057529    &1.149625859    &1.142380740    &$-$0.1319996114    &$-$0.1317252681 \\
&$2$ &23.77776210    &1.127432510    &1.117200151    &$-$0.1294445280    &$-$0.1291654762 \\
&$3$ &30.97134961    &1.106923322    &1.094073892    &$-$0.1270678978    &$-$0.1267825086 \\
&$4$ &37.01836535    &1.088034231    &1.072912395    &$-$0.1248619829    &$-$0.1245683079 \\
&$5$ &41.99388641    &1.070711790    &1.053639261    &$-$0.1228201001    &$-$0.1225157644 \\
&$6$ &45.96208086    &1.054912938    &1.036190429    &$-$0.1209365827    &$-$0.1206186343 \\
&$7$ &48.97706610    &1.040605011    &1.020513443    &$-$0.1192067694    &$-$0.1188714628 \\
&$8$ &51.08360344    &1.027766080    &1.006566764    &$-$0.1176270254    &$-$0.1172695033 \\
&$9$ &52.31764454    &1.016385716    &0.9943190061   &$-$0.1161947956    &$-$0.1158086001 \\
&$10$&52.70674195    &1.006466421    &0.9837479032   &$-$0.1149087585    &$-$0.1144850792 \\[1mm]
$L=2$
&$0$ &6.055104820    &1.171769108    &1.167629281    &$-$0.1345028469    &$-$0.1342319402 \\
&$1$ &15.75802764    &1.147923889    &1.140420315    &$-$0.1317734666    &$-$0.1314991895 \\
&$2$ &24.13150124    &1.125833576    &1.115363727    &$-$0.1292308991    &$-$0.1289518570 \\
&$3$ &31.27518973    &1.105423632    &1.092356539    &$-$0.1268663534    &$-$0.1265809080 \\
&$4$ &37.27560762    &1.086630566    &1.071309889    &$-$0.1246721507    &$-$0.1243783409 \\
&$5$ &42.20735955    &1.069401494    &1.052148054    &$-$0.1226416657    &$-$0.1223370994 \\
&$6$ &46.13417317    &1.053693917    &1.034807627    &$-$0.1207692907    &$-$0.1204509921 \\
&$7$ &49.10974929    &1.039475752    &1.019236796    &$-$0.1190504235    &$-$0.1187146132 \\
&$8$ &51.17844827    &1.026725684    &1.005394655    &$-$0.1174814900    &$-$0.1171232613 \\
&$9$ &52.37582778    &1.015433961    &0.9932504485   &$-$0.1160600096    &$-$0.1156728359 \\
&$10$&52.72904363    &1.005603867    &0.9827825290   &$-$0.1147847326    &$-$0.1143596945 \\[1mm]
$L=3$
&$0$ &6.742776308    &1.169077915    &1.164522384    &$-$0.1341475654    &$-$0.1338768278 \\
&$1$ &16.35924481    &1.145392790    &1.137506628    &$-$0.1314375645    &$-$0.1311633820 \\
&$2$ &24.65278168    &1.123456057    &1.112634818    &$-$0.1289136422    &$-$0.1286346099 \\
&$3$ &31.72221154    &1.103194065    &1.089805086    &$-$0.1265671000    &$-$0.1262815654 \\
&$4$ &37.65328342    &1.084544180    &1.068929626    &$-$0.1243903481    &$-$0.1240963309 \\
&$5$ &42.51989667    &1.067454360    &1.049933725    &$-$0.1223768484    &$-$0.1220719301 \\
&$6$ &46.38512106    &1.051882951    &1.032754956    &$-$0.1205210805    &$-$0.1202022490 \\
&$7$ &49.30203456    &1.037798741    &1.017342469    &$-$0.1188185314    &$-$0.1184819561 \\
&$8$ &51.31439742    &1.025181338    &1.003656310    &$-$0.1172657215    &$-$0.1169064227 \\
&$9$ &52.45717674    &1.014022013    &0.9916666688   &$-$0.1158602717    &$-$0.1154716137 \\
&$10$&52.75693246    &1.004325230    &0.9813528252   &$-$0.1146010528    &$-$0.1141739516 \\[1mm]
$L=4$
&$0$  &7.640242671   &1.165531695    &1.160431691    &$-$0.1336801802    &$-$0.1334096597 \\
&$1$  &17.14269529   &1.142058117    &1.133671151    &$-$0.1309957699    &$-$0.1307217054 \\
&$2$  &25.33081704   &1.120324362    &1.109043459    &$-$0.1284964706    &$-$0.1282174429 \\
&$3$  &32.30232202   &1.100257958    &1.086448221    &$-$0.1261737066    &$-$0.1258880439 \\
&$4$  &38.14194612   &1.081797407    &1.065799028    &$-$0.1240200062    &$-$0.1237257023 \\
&$5$  &42.92265458   &1.064891788    &1.047022508    &$-$0.1220289475    &$-$0.1217235484 \\
&$6$  &46.70664501   &1.049500571    &1.030057540    &$-$0.1201951261    &$-$0.1198755705 \\
&$7$  &49.54616878   &1.035593705    &1.014854552    &$-$0.1185141481    &$-$0.1181765359 \\
&$8$  &51.48418831   &1.023152039    &1.001374860    &$-$0.1169826603    &$-$0.1166219119 \\
&$9$  &52.55488390   &1.012168214    &0.9895899118   &$-$0.1155984217    &$-$0.1152077538 \\
&$10$ &52.78401937   &1.002648292    &0.9794802286   &$-$0.1143604633    &$-$0.1139305668 \\
\hline\hline
\end{tabular}
\end{center}
\end{table}

\section{Leading order radiative corrections}

The radiative corrections of an order $R_\infty\alpha^3$ for a one electron molecular system can be expressed by the following set of equations (see Refs.~\cite{Pac98,Yel01,HD_BL}). Only the spin-independent part is considered.

The one-loop self-energy correction ($R_\infty\alpha^3$):
\begin{equation}\label{se3}
E_{se}^{(3)} =
\frac{4\alpha^3}{3m_e^2}
         \left(
            \ln\frac{1}{\alpha^2}-\beta(L,v)+\frac{5}{6}-\frac{3}{8}
         \right)
         \left\langle
            Z_1\delta(\mathbf{r}_1)\!+\!Z_2\delta(\mathbf{r}_2)
         \right\rangle,
\end{equation}
where
\begin{equation}\label{Bethe}
\beta(L,v) =
   \frac{
   \left\langle
      \mathbf{J}(H_0\!-\!E_0)\ln\left((H_0\!-\!E_0)/R_\infty\right)\mathbf{J}
   \right\rangle}
   {\left\langle
      [\mathbf{J},[H_0,\,\mathbf{J}]]/2
   \right\rangle}
\end{equation}
is the Bethe logarithm. The latter quantity presents the most difficult numerical problem in computation of QED corrections for the three-body bound states. In \cite{KorobovBL12} the calculations for the ro-vibrational states in $\mbox{H}_2^+$ and $\mbox{HD}^+$ have been performed with sufficient precision. An operator $\mathbf{J}$ in (\ref{Bethe}) is the nonrelativistic electric current density operator of the system: $\mathbf{J}\!=\!\sum_a z_a\mathbf{p}_a/m_a$.

The anomalous magnetic moment ($R_\infty\alpha^3$):
\begin{equation}
E_{anom}^{(3)} = \frac{\pi\alpha^2}{m_e^2}
         \left[\frac{1}{2}\left(\frac{\alpha}{\pi}\right)\right]
         \left\langle
            Z_1\delta(\mathbf{r}_1)\!+\!Z_2\delta(\mathbf{r}_2)
         \right\rangle.
\end{equation}
Sometimes this term is incorporated into Eq.~(\ref{rc}) as a contribution from the form factors of an electron \cite{Kin96}.

The one-loop vacuum polarization ($R_\infty\alpha^3$):
\begin{equation}
E_{vp}^{(3)} = \frac{4\alpha^3}{3m^2}
         \left[-\frac{1}{5}\right]
         \left\langle
            Z_1\delta(\mathbf{r}_1)\!+\!Z_2\delta(\mathbf{r}_2)
         \right\rangle.
\end{equation}

The one transverse photon exchange ($R_\infty\alpha^3(m/M)$):
\begin{equation}\label{tr3}
E_{tr\text{-}ph}^{(3)}= \alpha^3\sum_{i=1,2}
    \left[
    \frac{2Z_i^2}{3m_eM_i}
        \left( -\ln\alpha-4\,\beta(L,v)+\frac{31}{3} \right)
        \left\langle \delta(\mathbf{r}_i) \right\rangle
    -\frac{14Z_i^2}{3m_eM_i}Q(r_i)
    \right],
\end{equation}
where $Q(r)$ is the $Q$-term introduced by Araki and Sucher \cite{as}:
\begin{equation}\label{Qterm}
Q(r) = \lim_{\rho \to 0} \left\langle
            \frac{\Theta(r - \rho)}{ 4\pi r^3 }
      + (\ln \rho + \gamma_E)\delta(\mathbf{r}) \right\rangle.
\end{equation}
It is worthy to note here that the splitting onto nonrecoil and recoil parts (Eqs.~(\ref{se3}) and (\ref{tr3})) is not exact, since the Bethe logarithm contains contributions both from the self-energy and exchange photon diagrams.

Summarizing the contributions one gets:
\begin{equation}
E_{\alpha^3} =
   E_{se}^{(3)}+E_{anom}^{(3)}+E_{vp}^{(3)}+E_{tr\text{-}ph}^{(3)}.
\end{equation}

\section{Results}

As it appears the mean value of the $\mathbf{p}_e^4$ operator is the most difficult quantity for numerical calculation. That is why we choose it for studying convergence of our results. Table \ref{converg} demonstrate convergence of this key quantity for the two vibrational states $v=4$ and $v=10$. The first case is needed to compare our numbers with previously obtained results by \cite{ZhongPRA12,ZhongCPB15}, in the latter case $v=4$ was the largest vibrational state considered. While in general there is a very good agreement between our data and the data from \cite{ZhongPRA12}, in case of the Hylleraas variational calculations it was observed that precision is going down rather rapidly with increase of the vibrational quantum number $v$. And already in case of the $(L\!=\!3,v\!=\!4)$ state we see some discrepancy, and we presume that it is primarily due to the difficulty in determining of the real uncertainty in the calculations. For the case of the vibrational state $v=10$ we were not able to get converged the last digit (tenth digit after the point). Still we may claim that for states of $v=9$ and $v=10$ at least nine digits after the point are significant and this precision is enough for theoretical predictions of vibrational transition energies at the level of $10^{-12}$ of relative uncertainty. As for other data in Tables \ref{H2plus-av} and \ref{HDplus-av} all the digits presented should be fixed.

Our main results of numerical calculations are presented in Tables \ref{H2plus-av} and \ref{HDplus-av}. The notation is as follows:
\begin{equation}
\begin{array}{@{}l}
\displaystyle
R_{ne} =
  -\left\langle
     \frac{\mathbf{p}_e\mathbf{P}_n}{r_n}
    +\frac{\mathbf{r}_n(\mathbf{r}_n\mathbf{p}_e)\mathbf{P}_n}{r_n^3}
  \right\rangle,
\qquad
R_{nn} =
  -\left\langle
     \frac{\mathbf{P}_1\mathbf{P}_2}{R}
    +\frac{\mathbf{R}(\mathbf{R}\mathbf{P}_1)\mathbf{P}_2}{R^3}
  \right\rangle,
\end{array}
\end{equation}
$Q_{ne}$ is the Q-term expectation value as it is defined in Eq.~(\ref{Qterm}), $n$ stands for one of the nuclei: $p$ or $d$. From these data one can easily get ro-vibrational transition intervals with account of the relativistic and radiative corrections of orders $R_\infty\alpha^2$, $R_\infty\alpha^2(m/M)$, $R_\infty\alpha^3$, and $R_\infty\alpha^3(m/M)$.

In summary, we have presented the extended systematic calculation of the leading order relativistic corrections for the ro-vibrational states of the hydrogen molecular ions $\mbox{H}_2^+$ and $\mbox{HD}^+$. It allows to comply with the present day requirements of the hydrogen molecular ion precision spectroscopy providing relative uncertainty due to numerical imperfection to the ro-vibrational transition frequencies below $10^{-12}$.

\section*{Acknowledgements}

The work was supported by the Ministry of Education and Science Republic of Kazakhstan under grant IRN AP05132978, V.I.K. acknowledges support of the "RUDN University Program 5-100" and funded by RFBR according to the research projects No. 12-34-56789 and No. 12-34-56789.

\end{document}